\documentclass[prb,showpacs,unsortedaddress,twocolumn,floatfix]{revtex4}
\usepackage[latin1]{inputenc}
\usepackage{color}
\usepackage{graphicx}
\usepackage{amsmath,amsfonts,amssymb}

\usepackage{graphicx,amssymb}

\begin{document}

\title{Evolution of unoccupied resonance during the synthesis of a silver
dimer on Ag(111)}
\author{A Sperl, J Kr\"{o}ger, and R Berndt}
\address{Institut f\"{u}r Experimentelle und Angewandte Physik,
Christian-Albrechts-Universit\"{a}t zu Kiel, D-24098 Kiel, Germany}
\author{A Franke and E Pehlke}
\address{Institut f\"{u}r Theoretische Physik und Astrophysik,
Christian-Albrechts-Universit\"{a}t zu Kiel, D-24098 Kiel, Germany}

\begin{abstract}
Silver dimers were fabricated on Ag(111) by single-atom manipulation using
the tip of a cryogenic scanning tunnelling microscope. An unoccupied electronic
resonance was observed to shift toward the Fermi level with decreasing
atom-atom distance as monitored by spatially resolved scanning tunnelling
spectroscopy. Density functional calculations were used to analyse the
experimental observations and revealed that the coupling between the
adsorbed atoms is predominantly direct rather than indirect via the Ag(111)
substrate.
\end{abstract}


\maketitle

\section{Introduction}
The electronic structure of adsorbed atoms (adatoms) or clusters of atoms
on surfaces determines the coupling between adsorbate and substrate
\cite{mpe_04,fol_04,lli05a,jkr_05}, the mutual interaction
between the adatoms \cite{nni_03}, magnetic properties \cite{jba_05}, as well
as their catalytic activity and selectivity \cite{sab_00,sab_01}. The coupling
between adatoms is particularly interesting since it plays a crucial role in
nucleation and is thus at the base of the microscopic understanding of thin
film growth on surfaces \cite{ble_78,jve_84,hbr_98}. Typically, the mutual
interaction comprises direct and indirect contributions. Direct interactions
result from the overlap of atomic orbitals and are responsible for the bonding
of dimers in vacuum. This type of interaction has been investigated for metal
dimers in the gas phase and noble-gas matrices \cite{sfe_93,hha_94,ira_99}.
An exponential energy splitting of bonding and antibonding states with the
atom-atom separation is characteristic for direct coupling. Indirect interactions
may become important for adatoms on surfaces and they depend strongly on the
electronic structure of the substrate. In particular, long-ranged and
oscillatory coupling between two adatoms or adsorbed molecules is mediated
by quasi-two-dimensional electronic continua
\cite{kla_78,mka_96,ewa_98,jre_00,nkn_02,fsi_04,tvh_06,mzi_08}.

The evolution of the electronic structure of clusters on surfaces with cluster
dimensions and geometric shapes has been analyzed atom by atom. Examples are
results from artificial gold chains \cite{nni_02},
quantum confinement of one-dimensional electronic states to chains of copper
atoms \cite{sfo_04}, unoccupied electronic resonances of silver clusters
with various sizes and shapes \cite{asp_08}, and the evolution of the Kondo
effect of a single magnetic atom with the number of adjacent non-magnetic
atoms in vertical \cite{nne_07} and lateral \cite{nne08b} hybridization
geometries. In a recent study of Au dimers on NiAl(110) a splitting of the Au
monomer resonance into bonding and antibonding states was reported as a function
of the Au-Au separation \cite{nni_03}. The emerging picture, which may be
inferred from this work, is that substrate-mediated adsorbate-adsorbate
interactions weaken the direct coupling between the adsorbates. In particular,
it was found in Ref.\,\cite{nni_03} that the splitting between bonding and
antibonding states varies linearly with the reciprocal mutual Au distance,
rather than exponentially as would be expected from a direct coupling in
vacuum.

Here, in a combined experimental and theoretical study, we investigated the
evolution of a Ag monomer $sp_z$ resonance, which shifts toward the Fermi
level upon approaching a second Ag atom. Our theoretical analysis indicated
that, at not too large adatom-adatom distances, the interaction between the
$sp_z$ electronic states on the surface is similar to the direct interaction
between the two $sp_z$ orbitals in vacuum.

\section{Experiment}
Measurements were performed with a custom-built scanning tunnelling microscope
operated at $7\,{\rm K}$ and in ultrahigh vacuum with a base pressure of
$10^{-9}\,{\rm Pa}$. The Ag(111) surface and chemically etched tungsten tips
were cleaned by argon ion bombardment and annealing. Individual silver atoms
were deposited onto the sample surface by controlled tip-surface contacts
as previously described in Ref.\,\cite{lli05b}. Using the tip of the
microscope, silver dimers were fabricated by atom manipulation. Spectra of the
differential conductance (${\rm d}I/{\rm d}V$) were acquired by superimposing
a sinusoidal voltage signal (root-mean-square amplitude $5\,{\rm mV}$, frequency
$4.7\,{\rm kHz}$) onto the tunnelling voltage and by measuring the current
response with a lock-in amplifier. Prior to spectroscopy of monomers and dimers
the tip status was monitored using spectra of the onset of the Ag(111) surface
state band edge. To obtain sharp onsets of the surface state signal and to
image single adatoms with nearly circular circumference the tip was controllably
indented into the substrate. Due to this {\it in vacuo} treatment the tip apex
was most likely covered with substrate material. All scanning tunnelling
microscopy (STM) images were acquired in the constant current mode with the
voltage applied to the sample. We divided the ${\rm d}I/{\rm d}V$ spectra by
$I/V$ to reduce the influence of the voltage-dependent transmission of the
tunnelling barrier \cite{jas_86}.

\section{Results and discussion}
\begin{figure}
  \includegraphics[width=85mm]{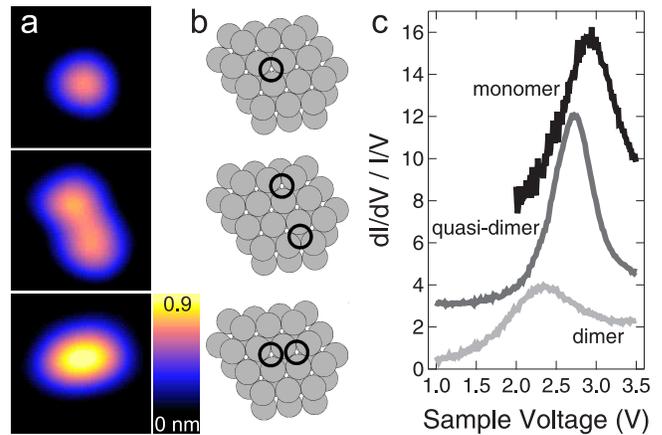}
  \caption[fig1]{(a) STM images of a single Ag adatom (top), two Ag adatoms
  with a mutual distance of $0.58\,{\rm nm}$, and Ag$_2$ on Ag(111) (sample
  voltage: $0.1\,{\rm V}$, tunnelling current: $0.1\,{\rm nA}$, image size:
  $2.1\,{\rm nm}\times 2.1\,{\rm nm}$). (b) Schematics of adsorption sites of
  silver adatoms (black circles) on Ag(111) lattice (gray dots). (c) Spectra
  of ${\rm d}I/{\rm d}V/(I/V)$ acquired with the tip placed above the centre
  of the assemblies in (a). The spectrum of the quasi-dimer and of the monomer
  have been shifted vertically by $3$ and $7$, respectively. Prior to
  spectroscopy the tip-sample distance had been set at $3.5\,{\rm V}$ and
  $1\,{\rm nA}$ for all spectra.}
  \label{fig1}
\end{figure}

Figure \ref{fig1}(a) presents a sequence of constant-current STM images
showing a single Ag adatom (top) on Ag(111), two Ag adatoms (middle) with a
distance of $\approx 0.58\,{\rm nm}$, and a silver dimer (bottom). Distances
between adatoms of Ag-Ag assemblies were determined from maxima positions of
cross-sectional profiles taken along the connecting line between the adatoms.
Together with the orientation of the assembly with respect to high symmetry
directions of the Ag(111) substrate, which was determined from dislocation
lines on the surface, the extracted adatom-adatom distances agree with lattice
site separations on Ag(111). The Ag dimer appears as a single entity in STM
images and we assigned the nearest-neighbour distance of Ag(111) to the
adatom-adatom distance of Ag$_2$. Our calculations indicated a slight
preference of the Ag adatom to occupy the face-centred cubic (fcc) adsorption
site to occupation of the metastable hexagonal close-packed (hcp) adsorption
site, which is of the order of the accuracy of the calculation.
We experienced that in the course of fabricating silver dimers individual
Ag adatoms were also found to occupy hcp adsorption sites [see the middle STM
image in Fig.\,\ref{fig1}(a) and the corresponding sketch in Fig.\,\ref{fig1}(b)].
Such assemblies occured frequently for adatom distances smaller than
$0.6\,{\rm nm}$. In the following these assemblies are referred to as quasi-dimers.
The schematics in Fig.\,\ref{fig1}(b) illustrate adsorption sites of individual
Ag adatoms (black circles) on the hosting Ag(111) lattice (gray dots) and the
orientation of the quasi-dimer and the dimer with respect to the Ag(111)
crystallographic directions. Figure \ref{fig1}(c) shows normalized
${\rm d}I/{\rm d}V$ spectra acquired on the centre of the single Ag adatom
(top), of a quasi-dimer (middle), and of the dimer (bottom). A gradual shift
of the monomer-related peak from $\approx 2.9\,{\rm eV}$ via a resonance energy
of $\approx 2.7\,{\rm eV}$ observed for the quasi-dimer to $\approx 2.3\,{\rm eV}$
for the dimer resonance binding energy was observed. The total shift of
$\approx 0.6\,{\rm eV}$ toward the Fermi level was not induced by the electric
field of the tip. Although the tip-surface distances may have differed for
the spectra shown in Fig.\,\ref{fig1}(c), it has been shown in
Refs.\,\cite{lli_03,jkr_04} that for shifts of the order of $10\,{\rm meV}$
the tip-surface distance had to be varied by several Angstroms, which was not
the case here.

\begin{figure}
  \includegraphics[width=85mm]{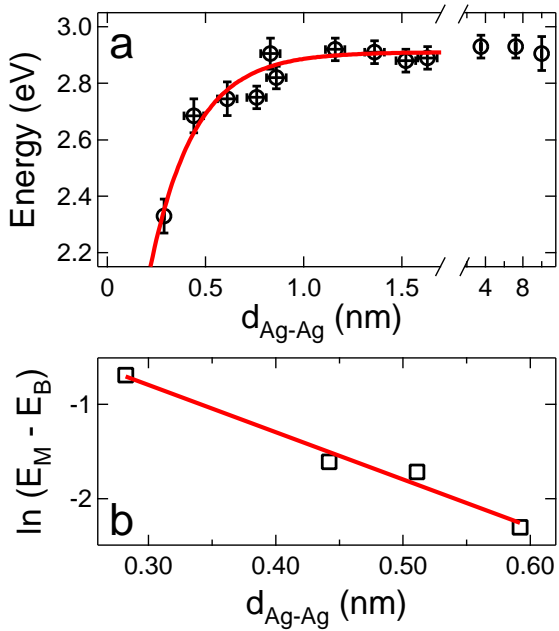}
  \caption[fig2]{(a) Energy of the unoccupied resonance as a function of the
  Ag adatom distance, $d_{{\rm Ag-Ag}}$. For distances larger than $2\,{\rm nm}$
  the spectra were acquired atop the individual Ag adatoms while for smaller
  distances spectroscopy was performed with the tip positioned above the centre
  of the assembly. The solid line is a fit to the calculated data in (b).
  (b) Logarithm of the calculated energy difference between monomer resonance
  (M) and bonding resonance (B) of an Ag-Ag assembly, $\ln(E_{\rm M}-E_{\rm B})$,
  plotted versus $d_{{\rm Ag-Ag}}$. The solid line is a linear fit to the data.}
  \label{fig2}
\end{figure}

Figure \ref{fig2}(a) summarizes the resonance energies measured for a variety
of Ag adatom separations, $d_{{\rm Ag-Ag}}$. For $d_{{\rm Ag-Ag}}>1\,{\rm nm}$
the shift of the resonance energy became too small to be resolved. For separations
$d_{{\rm Ag-Ag}}<1\,{\rm nm}$ a shift of the resonance energy became detectable
and increased rapidly as $d_{{\rm Ag-Ag}}$ approached the nearest-neighbour
distance of Ag(111). The solid line in Fig.\,\ref{fig2}(a) is a fit to calculated
data [Fig.\,\ref{fig2}(b)], which has been extrapolated to larger Ag-Ag
distances. The calculations are discussed below. For $d_{{\rm Ag-Ag}}>2\,{\rm nm}$
the spectra at the centre between two Ag adatoms were virtually identical with
spectra of clean Ag(111) in the relevant voltage interval \cite{asp_08}. For
this reason the data points for $d_{{\rm Ag-Ag}}>2\,{\rm nm}$ in Fig.\,\ref{fig2}
were acquired atop the individual adatoms.

Using density functional theory as implemented in the Vienna {\it ab initio}
Simulation Package (VASP) \cite{VASP1,VASP2,VASP3} we determined the electronic
structure of silver dimers adsorbed on Ag(111) with increasing adatom-adatom
separation. The generalized gradient approximation (GGA) PW91 by Perdew and
Wang \cite{PW_91} was applied to the exchange correlation functional. The
electron-ion interaction was treated within the framework of Bl\"{o}chl's
projector augmented wave (PAW) method \cite{PAW_94}. The potentials for VASP
were used from the database \cite{PAW2_99}. All configurations were modelled
in a slab geometry comprising of 14 layers of Ag. Silver dimers with separations
of $0.28$, $0.44$, $0.52$, and $0.59\,{\rm nm}$ were modelled in a
$(5\times 4)$, $(6\times 4)$, $(4\times 4)$, and $(6\times 4)$ surface unit
cell, respectively. The surface Brillouin zone was sampled with 16, 9 and 6
$k$ points for the $(4\times 4)$, $(5\times 4)$, and $(6\times 4)$ unit cells,
respectively. The Kohn-Sham wave functions were expanded in a plane wave basis
set with a $250\,{\rm eV}$ cutoff energy.

\begin{figure}
  \includegraphics[width=85mm]{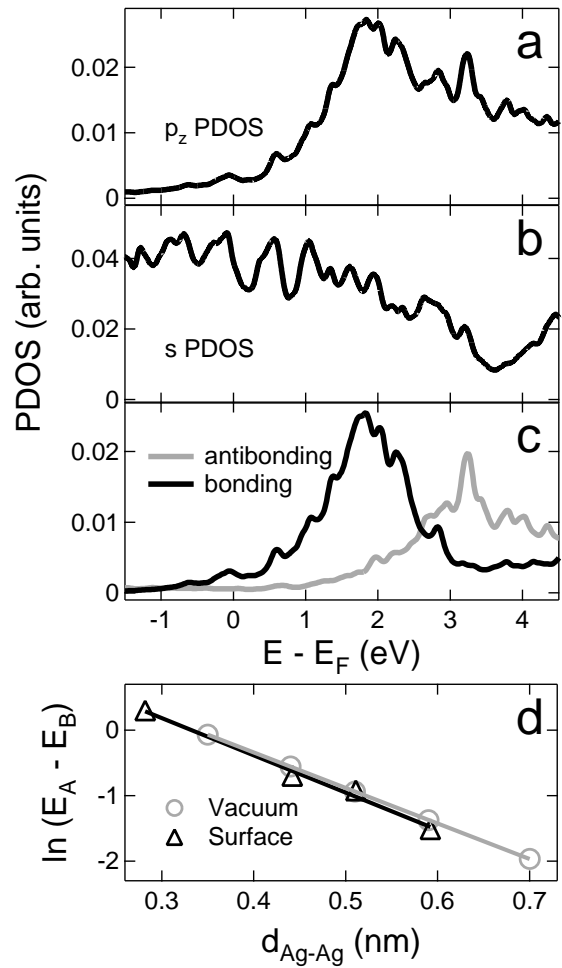}
  \caption[fig3]{(a) $p_z$ projected density of states (PDOS) and (b) $s$ PDOS
  for the silver dimer ($d_{{\rm Ag-Ag}} = 0.28\,{\rm nm}$, both adatoms at
  neighbouring fcc sites). (c) Density of states projected onto bonding and
  antibonding $p_z$ orbitals located at the two atoms of the Ag dimer.
  (d) Logarithm of the Kohn-Sham energy splitting between bonding (B) and
  antibonding (A) resonances as a function of Ag-Ag distance for Ag assemblies
  on the Ag(111) surface (triangles) and for free Ag atom pairs (circles)
  in the supercell. The solid lines are linear fits to the data.}
  \label{fig3}
\end{figure}

In order to identify the orbital composition of the experimentally observed
unoccupied resonance, we have calculated the projected density of states (PDOS)
with respect to atomic orbitals localized at the adatom sites. In case of the
silver dimer with both atoms at neighbouring fcc sites, the maximum of the
resonance was found between $1.8$ and $1.9\,{\rm eV}$ above the Fermi energy
\cite{asp_08}. This resonance can clearly be observed in the $p_z$ PDOS
[Fig.\,\ref{fig3}(a)], while it is absent in the $s$ PDOS [Fig.\,\ref{fig3}(b)].
From this we conclude a dominant $p_z$ character, with some $s$ admixture, of
this resonance.

By projecting onto bonding and antibonding combinations of $p_z$ orbitals
located at the two Ag atoms of the dimer, we identified the centre of the bonding
and antibonding $p_z$ resonance [Fig.\,\ref{fig3}(c)]. From this we obtained
the energy splitting between the bonding ($E_{{\rm B}}$) and the antibonding
($E_{{\rm A}}$) state, whose logarithm is plotted as a function of the mutual
Ag atom distance in Fig.\,\ref{fig3}(d). The results shown in Fig.\,\ref{fig3}(d)
are consistent with an exponential variation of $E_{{\rm A}}-E_{{\rm B}}$ within
the calculated range of Ag adatom distances. In order to compare to experiment,
we furthermore calculated the energy shift of the bonding resonance of the Ag
adatom pair with respect to the energy of the monomer resonance ($E_{\rm M}$)
[Fig.\,\ref{fig2}(b)]. We found that the energy splitting between the bonding
resonance with respect to the monomer resonance, $E_{\rm M}-E_{\rm B}$, quite
accurately coincides with half the energy shift between the bonding and the
antibonding states, $(E_{{\rm A}}-E_{{\rm B}})/2$. Thus, the splitting between
bonding and antibonding states is approximately symmetric. The solid line in
Fig.\,\ref{fig2}(a) represents a linear fit to $\ln(E_{\rm M}-E_{\rm B})$ as
a function of the Ag adatom separation $d_{{\rm Ag-Ag}}$, which has been
extrapolated to larger separations. We conclude that both experimental and
theoretical data are consistent with an exponential variation of the energy
splitting with $d_{{\rm Ag-Ag}}$. However, due to computational limitations
the range of Ag-Ag distances considered here for the two Ag adatoms on the
Ag substrate is too narrow to reliably distinguish between an exponential and
an algebraic ($1/d$) \cite{nni_03} dependence of energy splitting on adatom
separation.

For comparison we have calculated the energy splitting between the $5p_z$
bonding and antibonding orbitals of a Ag dimer in vacuum. The calculated
splitting of the silver $5p_z$ bonding and antibonding orbitals varied
exponentially with atom separation as expected \cite{sfe_93,ira_99}. Surprisingly,
as is evident from Fig.\,\ref{fig3}(d), the energy splitting between bonding
and antibonding states for the dimer in vacuum is similar to the splitting
of the dimer adsorbed on the surface for the Ag-Ag distances considered in
the {\it ab initio} calculations.

To further analyze the interaction between adatoms on a surface we resorted
to a minimum tight-binding model, which is similar to the model reported in
Ref.\,\cite{ks_98}. For simplicity we did not consider a tight-binding model
of the fcc crystal and its (111) surface, but we resorted to atoms with a
single $s$ orbital forming a simple cubic lattice, and two adatoms adsorbed
at on-top sites. We found a direct contribution, $2t$, to the energy
splitting between bonding and antibonding states, which is due to direct adatom
interaction. The parameter $t$ denotes the transfer integral for the two adatom
states, and $2t$ is identical to the splitting occuring for the free dimer
at the same adatom distance. Furthermore, there is a contribution to the
splitting owing to the interaction via the substrate, $2\,{\rm Re}(G_{ab})\,|v|^2$,
where $v$ denotes the next-neighbour transfer integral and $G_{ab}$ the matrix
element of the Green's function of the substrate with respect to the adsorption
sites. Calculations of such a matrix element of the Green's function for
two next-neighbour sites on the Ag(111) surface yielded values of
${\rm Re}(G_{ab})(\varepsilon)$ varying between $0.05$ and $0.15\,{\rm eV}^{-1}$
within the energy range of interest. For typical values of $|v|$ between $0.4$
and $1\,{\rm eV}$ the simple model also predicts that the interaction via the
substrate is distinctly smaller than the observed splitting. At large adatom
separations we expect the substrate-mediated interaction to eventually dominate
the direct interaction between the adatoms. In case of Ag atom pairs on Ag(111)
this appears to occur at a separation which is too large for the splitting to
be resolved in experiment.

In a previous investigation of Au dimers on NiAl(110) the splitting of bonding
and antibonding Au dimer states was reported for varying Au-Au distances
\cite{nni_03}. The authors of Ref.\,\cite{nni_03} found that the variation
of the splitting follows a $1/d_{{\rm Au-Au}}$ rather than an exponential
law. This observation was argued to be due to the influence of the substrate
electronic structure, which reduces the direct overlap of Au orbitals. In the
case of Ag-Ag assemblies on Ag(111), however, the interaction between the
individual Ag adatoms is similar to the one in vacuum [Fig.\,\ref{fig3}(d)].
A tentative explanation for this observation involves the electronic structures
of the substrates. The Ag(111) surface exhibits an extended $sp$ band
gap of surface-projected bulk electronic states in the centre of the surface
Brillouin zone \cite{ske_87}, while NiAl(110) does not \cite{slu_89,gca_92,zso_01}.
Since the resonance energy of the Ag-Ag assemblies falls into the surface-projected
$sp$ band gap it is likely that the Ag(111) substrate electronic structure
plays a less important role in mediating the interaction between the Ag adatoms.
As a consequence, the Ag adatom assemblies on Ag(111) are subject to a substrate
influence to a lesser extent than the Au adatom assemblies on NiAl(110). We
suggest that the large direct interaction in case of Ag dimers on Ag(111)
is also caused by the large extension of the $p$ wave function, which
dominates the resonance. This results in a slow decay of the transfer integral
as a function of Ag-Ag distance.

\section{Conclusion}
The interaction between two Ag adatoms on Ag(111) gives rise to a shift of a
$sp_z$ resonance toward the Fermi level with decreasing mutual adatom distances.
The shift was modelled by density functional calculations and is similar to
the shift calculated for a Ag dimer in vacuum. This observation indicates a
weak net influence of the substrate on the Ag-Ag interaction, which may originate
from the surface-projected $sp$ band gap of the substrate. We suggest that
adatom-adatom interactions on surfaces with band gaps in the relevant energy
interval exhibit a similar behaviour. As a further reason for the dominating
direct interaction we suggest a weak decay of atomic orbitals participating
in the interaction.

\section*{Acknowledgement}
Financial support by the Deutsche Forschungsgemeinschaft through SPP 1153 is
acknowledged.


\end{document}